%% file: cnt_soi.tex
\documentclass[aps,prl,reprint,superscriptaddress]{revtex4-1}
\usepackage{amsmath}
\usepackage[utf8]{inputenc}
\usepackage{lmodern}
\usepackage{siunitx}
\mathchardef\mhyphen="2D
\usepackage[final]{graphicx} 

\usepackage{adjustbox}

\usepackage{xcolor}

\makeatletter
\DeclareRobustCommand*\textsubscript[1]{%
  \@textsubscript{\selectfont#1}}
\def\@textsubscript#1{%
  {\m@th\ensuremath{_{\mbox{\fontsize\sf@size\z@#1}}}}}
\makeatother
\newcommand{\DeltaSO}{\ensuremath{\Delta_{\text{SO}}}}
\newcommand{\DeltaKK}{\ensuremath{\Delta_{KK'}}}
\newcommand{\sub}[1]{\textsubscript{#1}}
\newcommand{\bvec}[1]{\boldsymbol{\mathrm{#1}}}
\newcommand{\SOangle}[1]{\varphi_{\text{SO}}^{#1}}

\newcommand{\spinExp}{\left<\bvec{S}\right>}

\begin{document}


\title{Non-collinear spin-orbit magnetic fields in a carbon nanotube double quantum dot}

\author{M. C. Hels}
\affiliation{Center for Quantum Devices and Nano-Science Center, Niels Bohr Institute, University of Copenhagen, Universitetsparken 5, 2100~Copenhagen \O, Denmark}
\author{B. Braunecker}
\affiliation{SUPA, School of Physics and Astronomy, University of St.\ Andrews, North Haugh, St.\ Andrews KY16 9SS, United Kingdom}
\author{K. Grove-Rasmussen}
\author{J. Nyg\aa rd}
\affiliation{Center for Quantum Devices and Nano-Science Center, Niels Bohr Institute, University of Copenhagen, Universitetsparken 5, 2100~Copenhagen \O, Denmark}
\date{\today}

\begin{abstract}
    We demonstrate experimentally that non-collinear intrinsic spin-orbit magnetic fields can be realized in a curved carbon nanotube two-segment device.
    Each segment, analyzed in the quantum dot regime, shows near four-fold degenerate shell structure allowing for identification of the spin-orbit coupling and the angle between the two segments.
    Furthermore, we determine the four unique spin directions of the quantum states for specific shells and magnetic fields.
    This class of quantum dot systems is particularly interesting when combined with induced superconducting correlations as it may facilitate unconventional superconductivity and detection of Cooper pair entanglement.
    Our device comprises the necessary elements.
\end{abstract}
\maketitle

    Controlling local magnetic fields on the nanoscale is currently of strong interest in quantum information and spintronics.
    Local fields are important for operating spin quantum bits, spin-filters and realizing topological states of matter.
    Several approaches to address the spin degree of freedom in this manner have been pursued involving techniques based on on-chip strip lines \cite{KoppensNat2006}, nuclear spin ensembles \cite{PettaScience2005}, micromagnetic stray \cite{Pioro-LadriereNatPhys2008,KjaergaardPRB2014,Crisan2016} and exchange fields \cite{PasupathyScience2004,HauptmannNatPhys2008}, g-factor engineering \cite{SalisNat2001} and spin-orbit coupling \cite{Nowack2007,Nadj-Perge2010}.

    A particularly interesting situation arises in double dots coupled to a central superconducting electrode.
    Here the control of the local spin directions, inducing non-collinear spin projection axes for the two dots, is predicted to give rise to unconventional superconductivity \cite{SothmannPRB2014}, poor man's Majorana physics \cite{LeijnsePRB2012} and novel schemes for transport based spin-entanglement detection of Cooper pairs \cite{BrauneckerPRL2013}.

    These geometries can favorably be realized with carbon nanotubes when utilizing spin-orbit interactions (SOI) \cite{KuemmethNature2008,JespersenNatPhys2011,Klinovaja2011b,Bulaev2008,Izumida2009,Ando2000} to control \cite{PeiNatNano2012} and filter spins \cite{MazzaPRB2013}.
    In contrast to other quantum dot systems (e.g.\ semiconducting nanowires), the effect of SOI on the four-fold degenerate energy spectrum (due to spin $s=\uparrow, \downarrow$ and valley $\tau = K, K'$) of nanotubes is well-understood \cite{LairdRMP2015} and gives rise to spin-orbit magnetic fields oriented along the nanotube axis.
    In this Letter we demonstrate the realization of non-collinear magnetic fields using the spin-orbit fields in a bent carbon nanotube double quantum dot.
    The non-collinearity of the fields originates from the geometry-defined angle between the two quantum dot tube segments.
    Furthermore, the spin alignment can be controlled and rotated by applying an external magnetic field.

    \begin{figure}[b]
        \frame{\includegraphics[height=0.37\columnwidth]{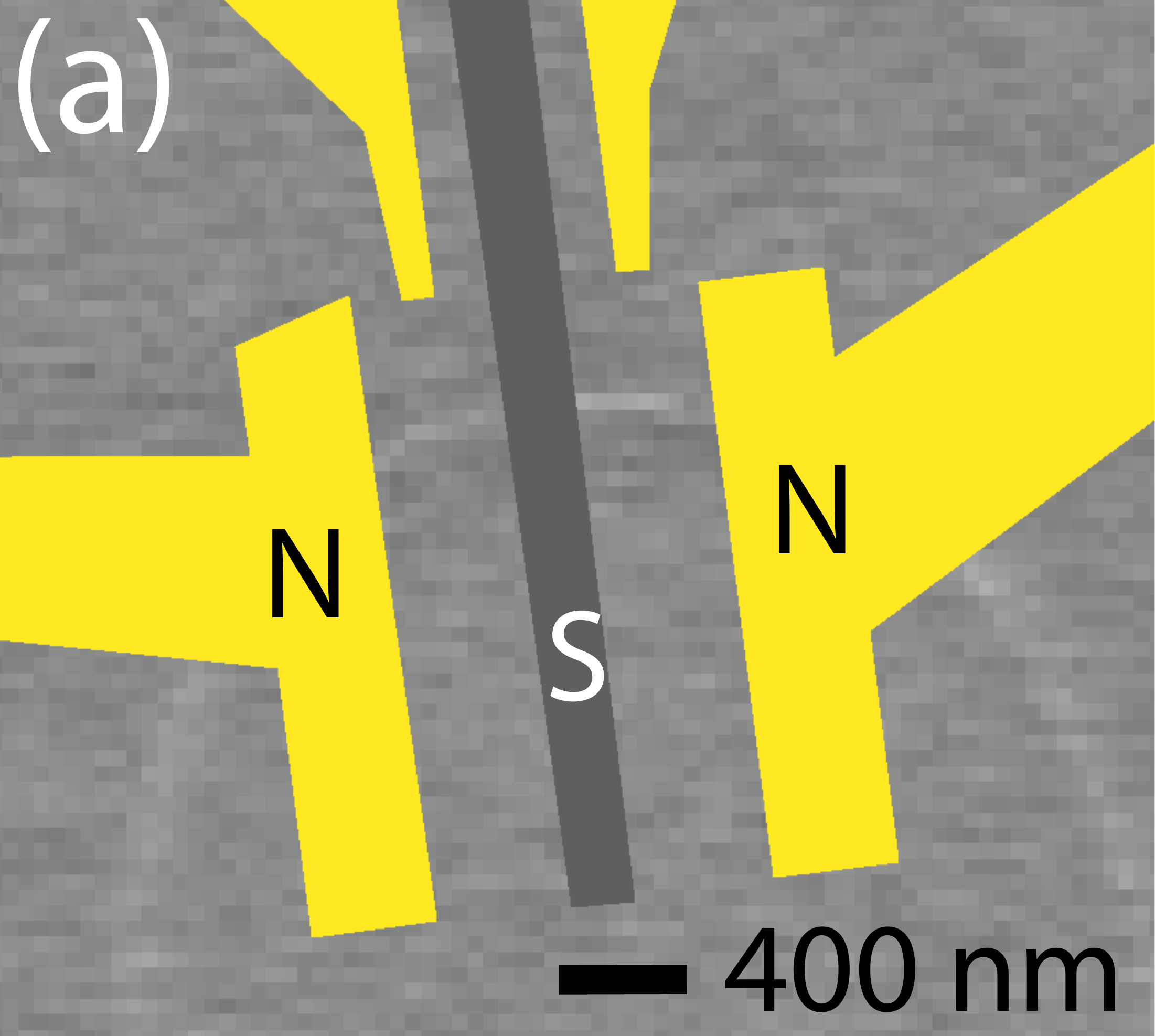}}%
        \hfill
        \frame{\includegraphics[height=0.37\columnwidth]{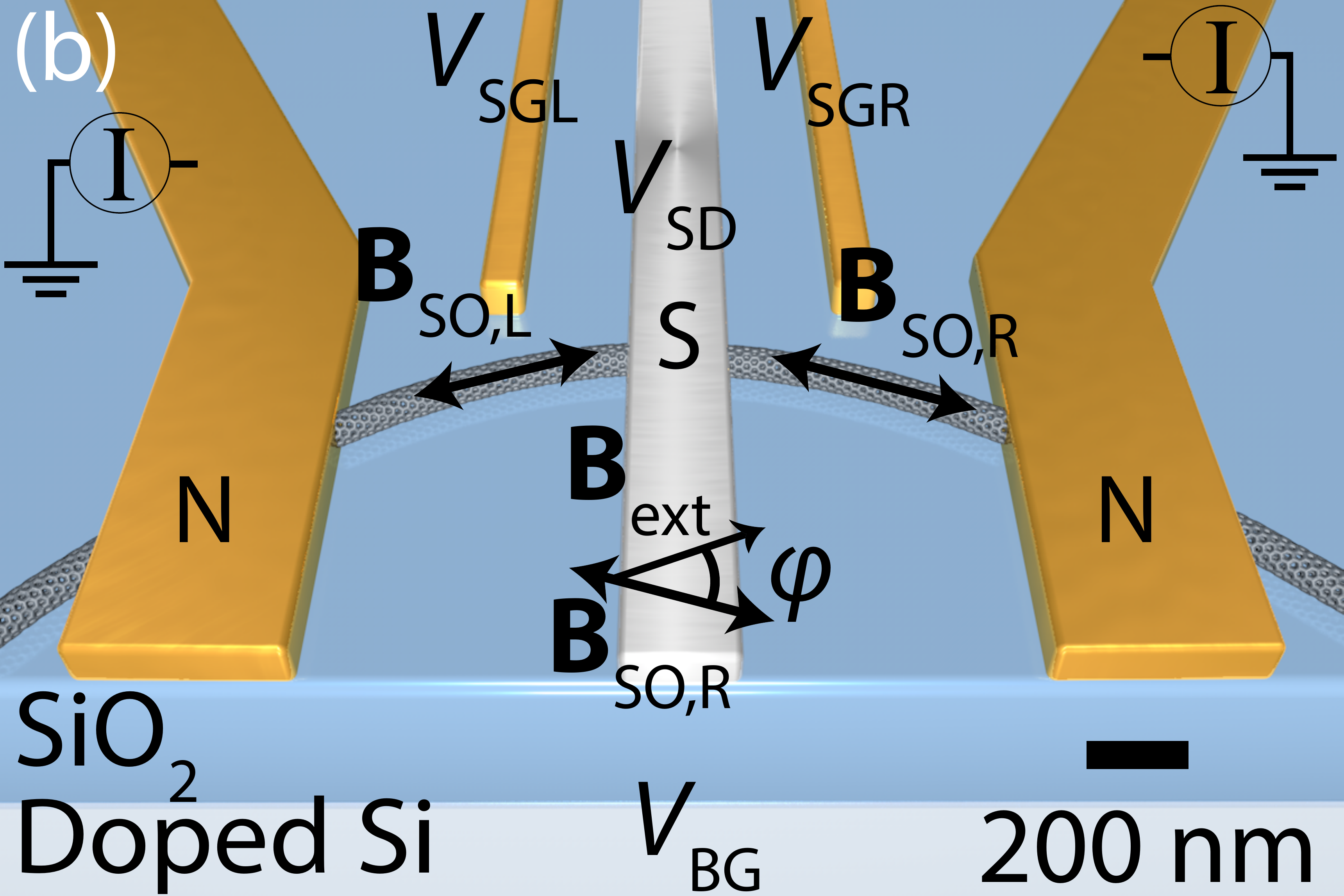}}%
        \caption{\label{fig:sem+schematic}
            \textbf{(a)} SEM image of nanotube with Au normal leads (\SI{50}{nm}), a Ti/Al superconducting lead (\SI{5/15}{nm}) and two side gates.
            Note that the lead geometries are overlaid on the image digitally.
            \textbf{(b)} Schematic of the device.
            The angle $\varphi$ reflects the angle between the external magnetic field $\bvec{B}\sub{ext}$ and the tangent of the midpoint of the right nanotube segment.
            A rigorous definition of $\varphi$ is given below.
            Note that $\bvec{B}\sub{ext}$ is constricted to the $x$-$z$-plane.
            The expected valley-dependent directions of the spin-orbit magnetic fields $\pm\bvec{B}\sub{SO,(L/R)}$ are indicated by the two-headed arrows.
        }
    \end{figure}
    In this paper we present data from the hybrid carbon nanotube device \cite{HerrmannPRL2010} shown in Fig.\ \ref{fig:sem+schematic}(a).
    Nanotubes are grown using chemical vapor deposition \cite{Kong1998} (CVD) on a doped Si chip with a 0.5\ $\mu$m thermal oxide cap.
    In the growth process nanotubes are bound to the substrate in random orientations by van der Waals forces.
    Many of the tubes are in this way fixated in curved configurations.
    The nanotubes are subsequently located using scanning electron microscopy (SEM) at an acceleration voltage of \SI{1.5}{kV} and metal leads are defined by e-beam lithography at \SI{20}{kV}.
    The two nanotube segments are \SI{450}{nm} long and the superconducting lead (\SI{5}{nm} Ti sticking layer with \SI{15}{nm} Al on top) is \SI{240}{nm} wide.

    The measurement configuration is indicated in Fig.\ \ref{fig:sem+schematic}(b).
    The superconducting lead is voltage biased and the normal leads are grounded through current amplifiers so that the nanotube segments are measured in parallel.
    Standard lock-in techniques and differentiated DC current were used to obtain $\mathrm{d}{I}/\mathrm{d}V$.
    The lock-in conductance was differentiated numerically to obtain $\mathrm{d}^2I/\mathrm{d}V^2$.
    Measurements were done at a base temperature of \SI{30}{mK}.

\begin{figure}[tb]
        \includegraphics[width=\columnwidth]{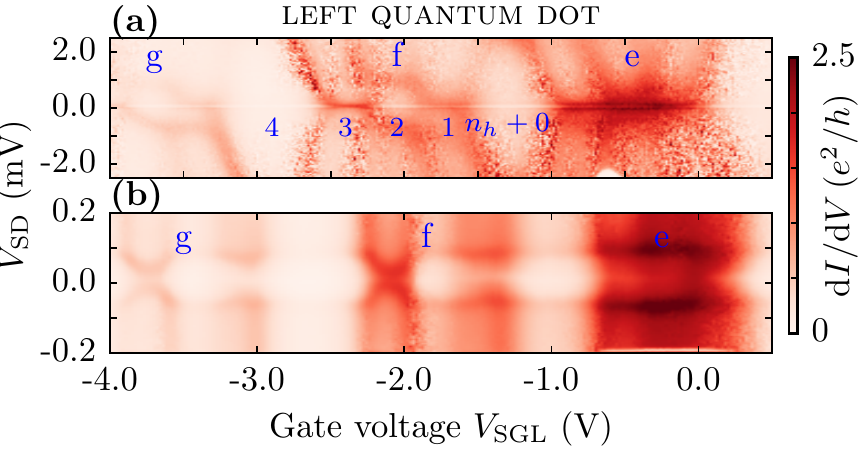}\\%
        \includegraphics[width=\columnwidth]{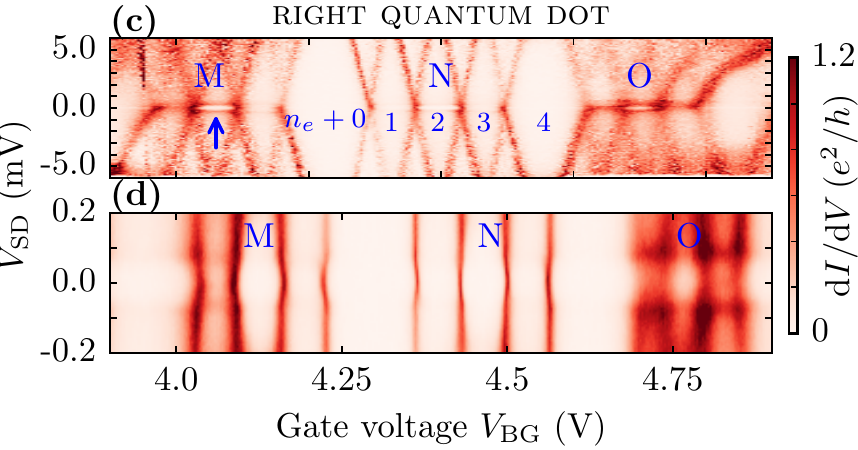}%
        \caption{\label{fig:bias_spec_gap_spec}
            \textbf{(a),(c)}
                Bias spectroscopy of the left and right quantum dots showing charging energies of $U\sub{L}\approx\SI{6}{meV}$ and $U\sub{R}\approx\SI{9}{meV}$, respectively.
                Both dots show four-fold shell filling characteristic of carbon nanotubes.
                Kondo resonances caused by strong lead-dot coupling are visible at zero bias for multiple fillings.
                They are most pronounced in shell e in the left dot.
                The blue arrow shows the the onset of inelastic cotunneling close to zero bias in shell M.
                The dot filling is $n_e=53$-$56$ electrons for shell N and $n_h=20$ holes for shell f.
            \textbf{(b),(d)}
                A superconducting gap is visible in both dots with a magnitude of about 60\ $\mu$eV.
        }
    \end{figure}

    Bias spectroscopy plots for the two nanotube segments are shown in Figs.\ \ref{fig:bias_spec_gap_spec}(a),(c).
    In both plots Coulomb resonances occur in sets of four reflecting the spin and valley degeneracy in carbon nanotubes at zero magnetic field \cite{LairdRMP2015}.
    Coulomb blockade in shell f in the left dot and shell M and O in the right dot is lifted by the SU(2) Kondo effect which is visible as zero-bias conductance resonances at electron fillings 1 and 3.
    In shell e in the left dot the lead-dot coupling is so strong that all four states participate in an SU(4) Kondo state which gives rise to a zero-bias conductance resonance for all fillings in the shell \cite{LairdRMP2015}.
    About 35 (60) shells are observed in the left (right) dot indicating that the nanotube is of high quality even though we do not employ an ultraclean fabrication procedure \cite{LairdRMP2015}.
    The right dot has a charging energy $U\sub{R}\approx\SI{9}{meV}$ and the left dot has a charging energy $U\sub{L}\approx\SI{6}{meV}$.
    In Fig.\ \ref{fig:bias_spec_gap_spec}(b),(d) bias spectroscopy plots at low $V\sub{SD}$ reveal a superconducting gap $\Delta\sub{SC}\approx 60\ \mu$eV.
    For $|V\sub{SD}|<\Delta\sub{SC}/|e|$ transport takes place via Andreev reflections.

    A narrow band gap is identified at $V\sub{BG}\approx\SI{0}{V}$ from room temperature data.
    We define the lettering of the shells so that lower (upper) case letters count in the direction of negative (positive) $V\sub{BG}$ starting with ``a'' (``A'') at the band gap.
    Thus, the filling in, e.g., shell N in the right side is 53-56 electrons within the uncertainty of the band gap position.

\begin{figure*}[tb]
        \includegraphics[width=\linewidth]{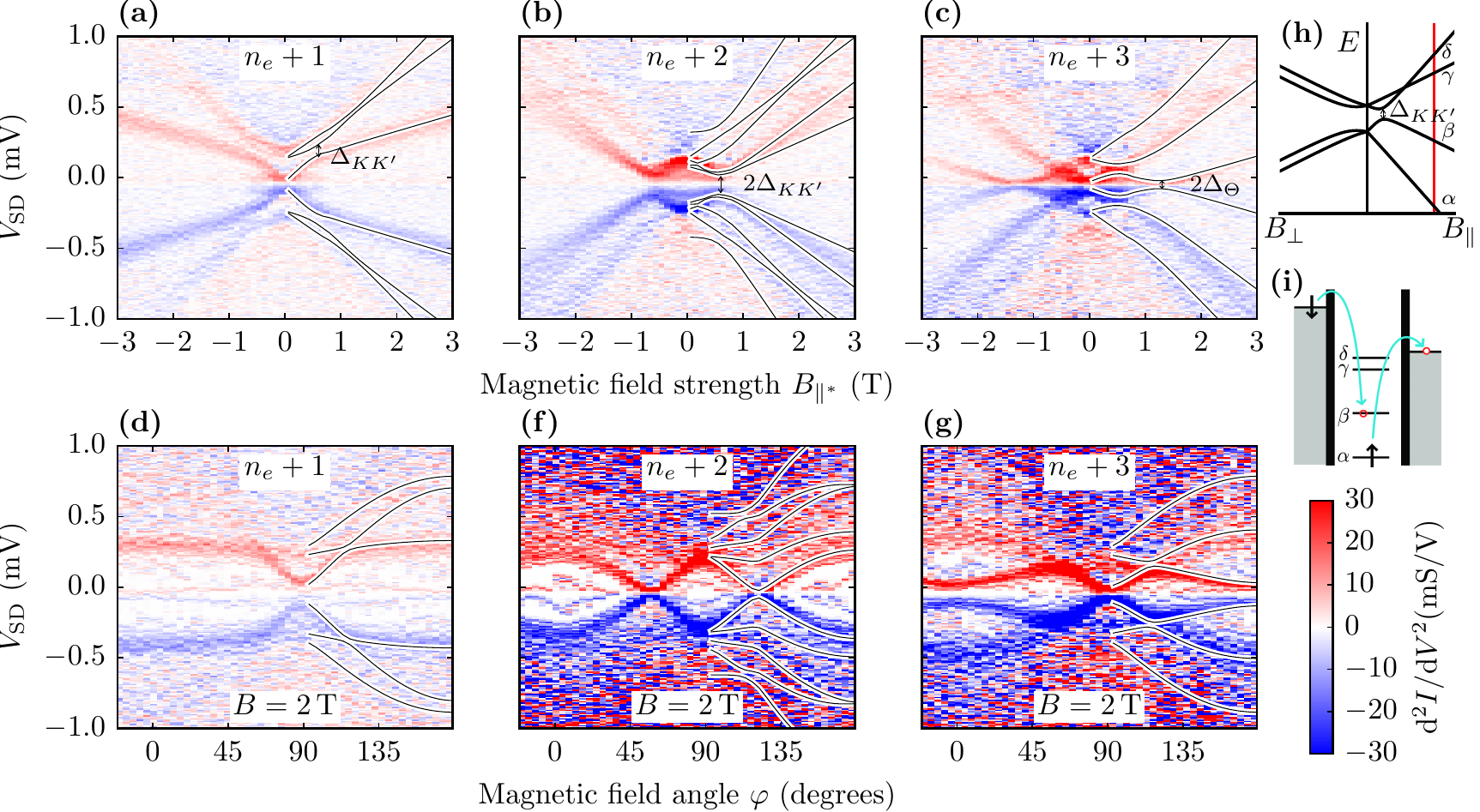}%
        \caption{\label{fig:ex_spec}
        Excitation spectroscopy of shell N in the right dot.
        \textbf{(a)-(c)}
            In the top row the magnetic field magnitude is swept with the field aligned $\SI{-11}{\degree}$ away from parallel orientation (denoted by $\parallel^*$).
            The misalignment does not change the spectrum qualitatively compared to parallel orientation.
        \textbf{(d)-(f)}
            Bottom row are magnetic field angle sweeps at $\SI{2}{T}$.
            The electron filling $n_e=53$-$56$.
            For all plots fitting parameters are $\DeltaSO = 150\ \mu \text{eV}, \DeltaKK = 70\ \mu \text{eV}, J = 120\ \mu \text{eV}$, $|g\sub{orb}|= 2.6$.
        \textbf{(h)}
            Plot of the nanotube spectrum with the parameters for the shell shown.
        \textbf{(i)}
            Schematic of an inelastic cotunneling process involving level $\alpha$ and $\beta$ at the red line in (h).
            This process is allowed when $|V\sub{SD}|$ is equal to or above the energy difference between levels $\alpha$ and $\beta$.
        }
    \end{figure*}

    We determine the electronic structure of a specific shell by inelastic cotunneling spectroscopy as sketched in Fig.\ \ref{fig:ex_spec}(i).
    When $eV\sub{SD}$ matches the energy difference between two levels in the nanotube there is an increase in conductance, i.e., a peak or a dip in $\mathrm{d}^2I/\mathrm{d}V_{\text{SD}}^2$.
    By varying the magnetic field strength and direction for different fillings (backgate voltages) in a shell the evolution of the four levels can be fitted to a four-level model \cite{LairdRMP2015,JespersenNatPhys2011,Klinovaja2011a,Klinovaja2011b,Bulaev2008,Weiss2010,Lim2011}.
    In this way we extract the spin-orbit energy $\Delta\sub{SO}$, valley coupling $\Delta_{KK'}$, orbital $g$-factor $g\sub{orb}$ and offset angle $\SOangle{}$ for the spin-orbit magnetic field which is parallel to the tube axis.

    SOI has the effect of lowering the energy of states with parallel spin and orbital magnetic moments.
    For the electron spin this is equivalent to the effect of a local magnetic field with opposite direction for the $K$ and $K'$ states.
    Thus, in zero magnetic field with only SOI present the four-fold degenerate states split into two doublets.
    Introducing disorder $\Delta_{KK'}$ couples $K$ and $K'$ states and opens avoided crossings.
    With both SOI and disorder present neither spin nor valley is a good quantum number.

    Fits for shell N are shown in Fig.\ \ref{fig:ex_spec} with excellent correspondence between spectroscopic data and model.
    In particular the electron-hole symmetry within the shell is seen to be broken by comparing (a) and (c) (see Fig.\ \ref{fig:ex_spec}(h)).
    For this shell we find $\DeltaSO=120\mhyphen170\ \mu\text{eV}, \DeltaKK=50\mhyphen100\ \mu\text{eV}, |g\sub{orb}|=2.4\mhyphen2.9$ and exchange splitting $J=50\mhyphen200\ \mu\text{eV}$, which are consistent with values previously reported \cite{KuemmethNature2008,Churchill2009,JespersenNatPhys2011,Lai2014,Cleuziou2013,Schmid2015}.
    Some transitions are not visible in the magnetic field angle sweeps because of noise.
    The offset angle for shell N, $\SOangle{\text{N}}$ is chosen as the zero-point of $\varphi$ so that $\SOangle{\text{N}}\equiv 0$.
    This angle is found to coincide with the tangent of the midpoint of the right segment in the SEM image within experimental uncertainty.

    Interestingly, shell N is spin-orbit dominated by having $\Delta\sub{SO} > \Delta_{KK'}$.
    A nanotube in this regime is not normally obtained in devices that are not ultraclean since exposing the nanotube to lithographic processing typically introduces disorder.
    The property of being spin-orbit dominated shows up clearly in the difference between Fig.\ \ref{fig:ex_spec}(a) and (c) which would be identical \cite{JespersenNatPhys2011} for $\Delta\sub{SO}=0$.
    With SOI the two highest-energy levels cross at the spin-orbit magnetic field $B\sub{SO} \equiv \Delta\sub{SO}/g\sub{s}\mu\sub{B}$.
    Disorder does not cause an anti-crossing here since the two states belong to the same valley.
    However, the spin-flip scattering by an external magnetic field $\bvec{B}\sub{ext}$ making an angle $\Theta$ with $\bvec{B}\sub{SO}$ causes an anti-crossing of $\Delta_{\Theta}$ \cite{Churchill2009} which is observed in Fig.\ \ref{fig:ex_spec}(c).

    Of particular interest is the angle $\SOangle{\nu}$ between the directions of the intrinsic spin-orbit magnetic field in a given shell $\nu$ and N.
    The $\SOangle{\nu}$ are extracted from magnetic field angle sweeps by identifying the angle with which the fit must be offset for it to correspond to the data.
    The model does not take into account the curving of the tube and thus $\SOangle{\nu}$ represents an effective angle dependent on the position of the wave function of each shell $\nu$.
    This interpretation implies that $\SOangle{\nu}$ is the same for levels in the same shell, which is consistent with the data.

\begin{figure}[tb]
        \includegraphics[width=\linewidth]{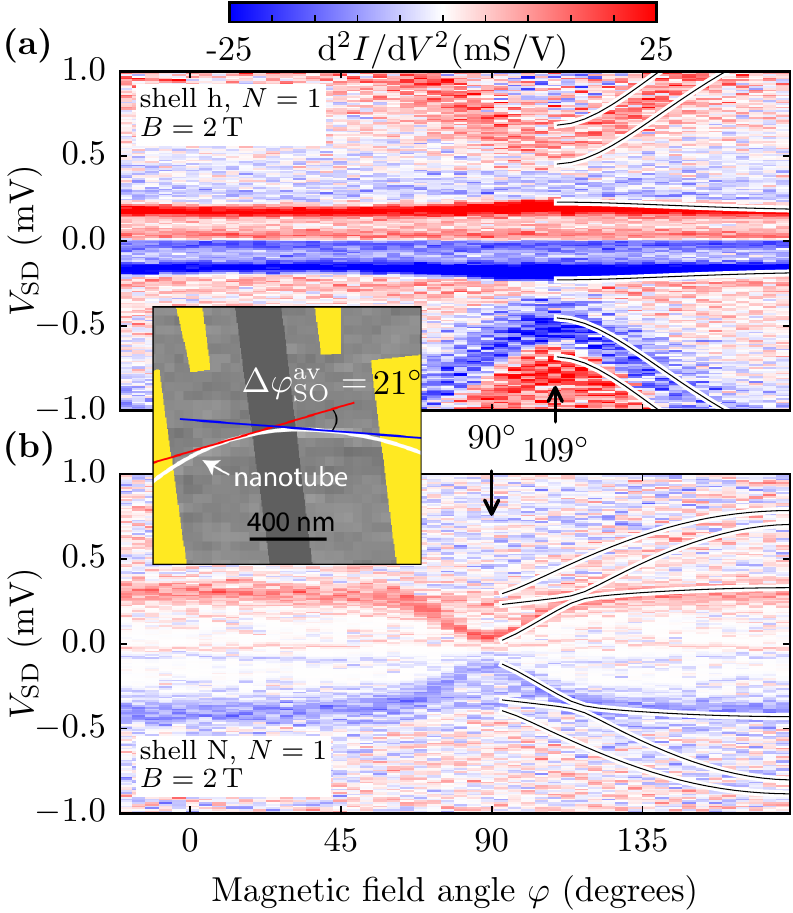}\\
        \includegraphics[width=\linewidth]{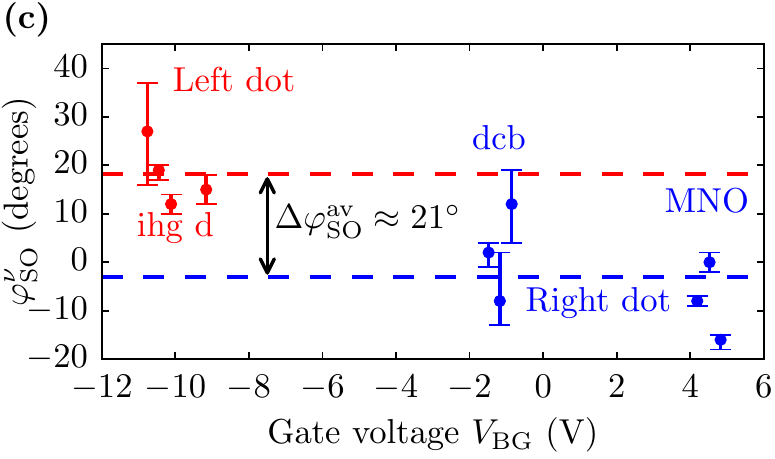}
        \caption{\label{fig:angle_comp_scatter}
            Excitation spectroscopy of \textbf{(a)} shell h in the left segment and \textbf{(b)} shell N in the right segment.
            Annotations show the angle where $\bvec{B}\sub{ext}$ is oriented perpendicular to the spin-orbit magnetic field.
            This occurs for different angles in the two segments.
            \textbf{Inset}:
                SEM image of the device.
                The average angle difference between the left and right dot of \ang{21} is illustrated by tangents drawn on the nanotube segments.
                Drawing the tangents at the segment midpoints would instead give $\Delta\SOangle{\text{av}}\approx\ang{32}$.
            \textbf{(c)}
                Spin-orbit magnetic field angles plotted for the shells measured.
                Red (blue) denotes left (right) segment.
                Dashed lines are averages of angles in a segment.
                Error bars are obtained as the minimum and maximum value for $\SOangle{\nu}$ that make the model fit the data.
        }
    \end{figure}

    We now want to establish that the $\SOangle{\nu}$ are, indeed, different in the two nanotube segments.
    Figure\ \ref{fig:angle_comp_scatter} shows a comparison between excitation spectroscopy for shell h in the left segment and shell N in the right segment (shell N is also shown in Fig.\ \ref{fig:ex_spec}).
    In this figure the minimum of the second (first) transition for shell h (N) can be identified as the angle where the external magnetic field is oriented perpendicular to the spin-orbit magnetic field.
    The minimum occurs because the magnetic field couples minimally with the orbital magnetic moment in the nanotube at perpendicular orientation.
    We find that the spin-orbit magnetic field angles $\SOangle{\text{h}}=\ang{109}-\ang{90}=\ang{19}$ and (by definition) $\SOangle{\text{N}}=\ang{0}$ are at an angle of \ang{19} with respect to each other.

    \begin{figure}[tb]
        \includegraphics[width=\linewidth]{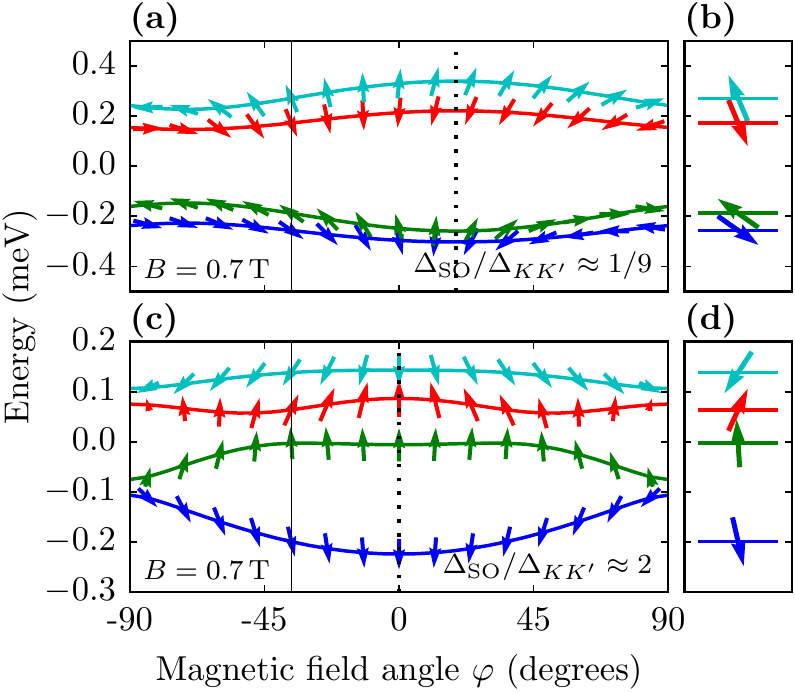}
        \caption{\label{fig:spin_direction}
            \textbf{(a),(c)} Spectrum and spin expectation value $\left<\bvec{S}\right>$ as a function of external magnetic field angle $\varphi$.
            \textbf{(b),(d)} Energy levels and $\left<\bvec{S}\right>$ at the position of the solid black lines in (a),(c).
            The model parameters are for shell h and shell N in (a),(b) and (c),(d), respectively.
            The spin vectors are shown in the $x$-$z$-plane since the effective magnetic field is constricted to this plane.
            The dotted vertical lines show the angle at which the magnetic field is parallel to the tube segment.
            All spins are shown in a global basis relative to shell N.
        }
    \end{figure}

    Similar sets of measurements of parallel, perpendicular and angular magnetic field sweeps have been made for nine other shells.
    The extracted angles $\SOangle{\nu}$ are plotted in Fig.\ \ref{fig:angle_comp_scatter} in which a clear clustering is seen with the $\SOangle{}$-angles generally being smaller in the right dot.
    From the $\SOangle{\nu}$-values we estimate the average angle for the left and right segment $\SOangle{\text{av,(L,R)}}$ and the difference between the averages $\Delta\SOangle{\text{av}}=\SOangle{\text{av,L}} - \SOangle{\text{av,R}}=\ang{21}\pm\ang{3}$.
    This angle is visualized in the inset of Fig.\ \ref{fig:angle_comp_scatter} by tangents drawn on the nanotube segments.
    The large uncertainties in shells i, c and b are due to low $g\sub{orb}$-values which reduces the curvature in the excitation spectroscopy plots \footnote{
        The data set consists of 77 sweeps similar to the six sweeps in Fig.\ \ref{fig:ex_spec}.
        Only for the $n_e=3$ sweep for shell d (i.e., one sweep out of 77) is the value for $\SOangle{}$ inconsistent with the other sweeps for the shell.
        The correspondence between data and model for this particular sweep is not as good as for the remaining sweeps in this shell, even if the angle is chosen to fit only this shell.
        One explanation is that the longitudinal wave function for this filling is different from the other fillings.
    }.

    The present device geometry allows for injection of Cooper pairs from the central superconducting electrode into the two dots.
    Ideally, as one electron tunnels into each dot, the Cooper pair is split into the two nanotube segments \cite{Hofstetter2009,HerrmannPRL2010}, thereby injecting non-local spin-entangled electrons in the two dots \cite{Recher2001}.
    This type of device may allow tests of a Bell inequality for Cooper pairs \footnote{In such an experiment, the violation of a Bell inequality is not used to disprove local
    hidden variable theories \cite{Bell1964}, but as a proof of entanglement under the assumption that quantum mechanics is valid.}.
    Notably, it has recently been proposed that with the bent nanotube geometry the effective magnetic field due to SOI and external field, can be used to configure levels with two spin projection axes per segment to act as spin filters for entanglement detection \cite{BrauneckerPRL2013,MazzaPRB2013,Burset2011}.
    This scheme therefore requires a clear quantum dot shell structure, SOI, and a finite angle $\SOangle{}$ between the spin-orbit fields for the two dots as shown above.

    To address the spin projection axes we show in Fig.\ \ref{fig:spin_direction}(a),(c) the spin expectation value $\spinExp$ for shells h and N as a function of external magnetic field angle $\varphi$ based on the four-level model.
    For $\DeltaKK=0$, all quantum dot levels are fully spin polarized, with
    directions of $\spinExp$ that are different in the two quantum dots because of different $\SOangle{}$ and $\Delta\sub{SO}$.
    In reality $\DeltaKK$ is always present, and the spin polarization is only partial.
    Nonetheless, as long as $\DeltaKK$ remains smaller than the minimum energy difference $\delta E$ between
    the levels with opposite valley in the same shell (see Supplemental Information), a detection of entanglement remains possible
    upon a prudent selection of angular ranges and magnetic field strengths, for which the polarizations remain close to full,
    e.g., as indicated in Fig. \ref{fig:spin_direction}(b),(d) at $\varphi=\ang{-36}$.
    For systems as our sample, $\DeltaKK$ should lie in the
    sub 100~$\mu$eV range to fulfill this requirement, which puts shell N in the good regime but would require further
    improvement for shell h.
    While challenging, this is within experimental reach and shows that our device is close to fulfilling the conditions for an entanglement test.
    Furthermore, our theoretical understanding for $\DeltaKK\neq 0$ (Supplemental Information) allows analysis even in the non-ideal regime.

    In conclusion, we have shown that non-collinear spin-orbit magnetic fields with a significant average angle of over \ang{20} degrees can be realized in a curved nanotube double dot device.
    The underlying analysis of fits of the energy spectrum versus magnetic field allows for precise predictions of the spin orientation in the two quantum dots.
    Spin control has implications for several proposals coupling quantum dots to superconductors \cite{LeijnsePRB2012,SothmannPRB2014,Shekter2016}.
    In particular, the distinctly non-collinear spin projection axes can be used for spin detection and future testing of spin-entanglement of split Cooper pairs \cite{BrauneckerPRL2013}.

    \begin{acknowledgments}
        We thank A. Levy~Yeyati, and J. Paaske for fruitful discussions and acknowledge the financial support from the Carlsberg Foundation, the European Commission FP7 project SE2ND, the Danish Research Councils and the Danish National Research Foundation.
    \end{acknowledgments}

\input{cntsoistatic.bbl}

\end{document}

%% file: cntsoistatic.bbl
%